\documentclass{svproc}
\usepackage{graphicx}
\usepackage{amsmath}
\usepackage{amssymb}
\usepackage{dsfont}
\usepackage{xcolor}
\usepackage{upgreek}
\usepackage{enumerate}
\usepackage{tikz}
\usepackage{amsfonts}
\usepackage{textcomp}
\usepackage[normalem]{ulem}
\usepackage{setspace}
\usepackage{comment}
\usepackage{relsize}
\usepackage{subfig}
\usepackage{cite}

\usetikzlibrary{shapes,arrows}
\usepackage{verbatim}

\usepackage{arydshln,leftidx,mathtools}
\usepackage{caption}
\usepackage{float}
\usepackage{bm}
\newtheorem{defn}{Definition}

\newtheorem{coro}{Corollary}
\newtheorem{assumption}{Assumption}

\newcommand{\norm}[1]{\left\lVert#1\right\rVert}

\DeclareMathOperator{\vect}{vec}

\begin{document}
\mainmatter              
\title{Mean-Square Stability And Stabilizability Analyses of LTI  Systems Under Spatially Correlated  Multiplicative Perturbations}
\titlerunning{Mean-Square Stabilizability}  
%
\author{Jianqi Chen\inst{1}\and Yanling Ding\inst{1} \and Hui Peng\inst{2} \and Tian Qi\inst{3}  \and Jie Chen\inst{1}}
\authorrunning{Jianqi Chen et al.} 
%
\tocauthor{Jie Chen}
\institute{Department of Electrical Engineering,  City University of Hong Kong,  Hong Kong,  China.\\
\email{jianqchen2-c@my.cityu.edu.hk; ylding4-c@my.cityu.edu.hk; jichen@cityu.edu.hk}
\and
 School of Automation,  Guangdong University of Technology,  Guangzhou,  China \\
\email{penghui0816@163.com}
\and
School of Automation Science and Engineering,  South China University of Technology,  Guangzhou,  China.\\
\email{auqt@scut.edu.cn}}

\maketitle              

\begin{abstract}
In this paper, we first study the robust stability problem for discrete-time linear time-invariant systems under  stochastic 
multiplicative uncertainties. Those uncertainties could be are susceptible to describing  transmission errors, packet drops, random delays, and fading phenomena
in networked control systems. In its full generality, we assume that the multiplicative uncertainties are diagonally structured, and 
are allowed to be spatially correlated across different patterns, which differs from previously related work significantly.
We derive a necessary and sufficient condition for robust stability in the mean-square sense against such uncertainties.  
Based on the obtained stability condition, we further investigate the mean-square stabilizability and consensusability problems through
two case studies of first-order single- and two-agent systems. The  necessary and sufficient conditions to guarantee stabilizability and consensusability are derived, which
rely on the unstable system dynamics, and the stochastic uncertainty variances.

\end{abstract}
\section{Introduction}
Networked control is widely considered a key enabling and transformative technology for the current- and next-generation engineering systems. 
Networked control systems differ from conventional feedback systems, in which networks are implemented to perform communications and enable exchange of data.
For such systems, as the most important feature, multiple tasks can be executed remotely combining the cyberspace and physical space, which reduces effectively the complexity and the overall cost in designing and implementing the control systems. 
However, the consequent communication noises and transmission losses over uncertain networks usher in new challenges inevitably. 
Over the past decade, the stability and performance problems of networked control systems have receive compelling attention from the control community
(see, e.g.,\cite{hinrichsen1996stability,brockett2000quantized,elia2005remote,lu2002mean,braslavsky2007feedback,schenato2007foundations,bamieh2012structured,
qi2017control,su2017control,chen2018mean}).
Recent studies in \cite{elia2005remote,chen2018mean} reveal that stochastic multiplicative noises can be used to effectively model network communication uncertainties including the random-delay and  data-loss  network phenomena. 
This relevance of multiplicative channel noises to networked control systems results in a direct impetus motivating our study.

Throughout of this paper, the channel uncertainties are modelled as diagonally structured multiplicative perturbations, which consist of static, zero-mean
stochastic processes. The current model and theory of networked control, however,  are unable to break through one fundamental limitation, that is the communication channel noises must be independent or uncorrelated \cite{elia2005remote,lu2002mean,you2010minimum,qi2017control}. In this paper, we concentrate on the  correlated stochastic multiplicative uncertainties coping with correlated noises, transmission losses, and a wide range of other channel models.
Under this formulation, we may enable to construct a framework to examine the robust stability and performance problems of networked control systems under such uncertainties. 
Note that the descriptions of uncertainty are different from those in robust control theory \cite{zhou1996robust}.  In doing so, we assess the system's stability and performance
using \textit{mean-square measures} \cite{lu2002mean,qi2017control}.

In this paper, we focus on the mean-square stability and stabilizability problems. 
In the presence of correlated stochastic multiplicative uncertainties,
we seek to develop
fundamental necessary and sufficient conditions that guarantee the stabilizability of linear time-invariant (LTI) systems by the output feedback controller in the mean-square sense.
We first develop a mean-square stability condition, namely a
generalized mean-square small gain theorem capable of coping with correlated stochastic uncertainties.  
Next, under the obtained generalized mean-square framework, we attempt to solve the corresponding mean-square stabilizability problems. At current stage, a complete solution to the generalized  LTI systems are still unavailable. We consider two special but representative low-order systems serving as a case study. The first case is the single-input single-output (SISO) first-order unstable plant. The channels between the plant and the controller are perturbed by correlated stochastic uncertainties. We develop the necessary and sufficient mean-square stabilizability conditions for such plant. We next consider a typical multi-agent system containing only two agents, each of which keeps a first-order dynamics. Then 
the associated mean-square consensusability condition is derived, linking the unstable pole  of the agent and the variances of the uncertainties together.

The notations used throughout of this paper are collected herein. 
Let $\mathbb{R}^{n}$ and $\mathbb{R}^{m\times n}$ 
be the space of real vectors and 
real matrices. For any matrix $A\in \mathbb{R}^{n\times n}$, we denote  $\rho(A)$, 
$A^{\top}$, $A^{\ast}$, $A^{\mathrm {H} }$, $A_{ij}$, and $\vect{(A)}$ as  the spectral radius, the transpose, the conjugate, the conjugate and transpose, the $ij$th entry, and the column stack, respectively. 
Given two matrices $A,~B\in\mathbb{R}^{n\times n}$,  $A\leq (\geq) B$ implies the Loewner order.
$I_n\in \mathbb{R}^{n\times n}$ denotes a unit matrix, and we may omit the subscript $n$ if the dimension is apparent
in context. $\textbf{1}_n\in \mathbb{R}^{n}$ represents the vector with all entries equal to one.
$\otimes$ and $\circ$ denote the Kronecker product and Hadamard product.
We denote the expectation operator by 
$\mathbb{E}[\cdot]$. 
 At last, given a LTI stable system $G$, we use $\norm{G}_2$ to represent its $\mathcal{H}_2$ norm.  

\section{Problem Formulation and Preliminaries} 
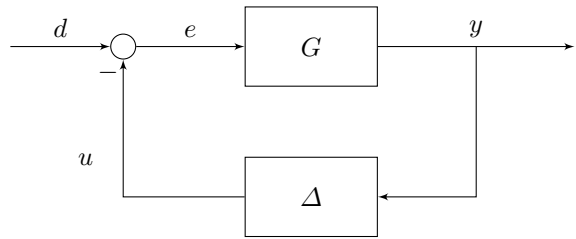
\begin{figure}
\begin{center}
\begin{tikzpicture}[auto, node distance=3cm,>=latex']
\tikzstyle{block} = [draw, rectangle, 
    minimum height=3em, minimum width=5em]
\tikzstyle{sum} = [draw, circle, node distance=1cm]
\tikzstyle{input} = [coordinate]
\tikzstyle{output} = [coordinate]
\tikzstyle{pinstyle} = [pin edge={to-,thin,black}]

   \node [input, name=input] at (-0.5,0) {};
   \node [sum] (sum) at (1,0) {};
   \node [block] (plant) at (3.5,0) {$G$};
   \node [block] (uncertainty) at (3.5,-2)  {$\Delta$};
   \node [output, name=output]  at (7,0) {};       
            
  \draw [->] (input) -- node[name=d] {$d$} (sum);    
  \draw [->] (sum) -- node[name=e] {$e$} (plant);
  \draw [->] (plant) -- node[name=y] {$y$} (output);
  \draw [->] (y) |- (uncertainty);
  \draw [->] (uncertainty) -| (sum);
   \node  at (0.5,-1.5) {$u$};
   \node  at (0.8,-0.35) {$-$};
\end{tikzpicture}
\end{center}
\caption{The LTI system with the stochastic multiplicative uncertainty.}
\label{interconnection}
\end{figure}

For consideration of stability and stabilization problems, we start by the
uncertain system depicted in Fig. \ref{interconnection}. In this configuration, 
$G$ represents an open-loop stable discrete-time LTI plant. 
The uncertainty $\Delta$ here 
admits a diagonal structure such that
\begin{equation}\label{eq:delta}
\Delta(k)=\mathrm{diag}\left( \Delta_1(k),\ldots,\Delta_m(k) \right),
\end{equation}  
and
 $$u(k)=\Delta(k)y(k).$$
Note that each component $\Delta_i(k),~i=1,\ldots,m$ is a zero-mean stochastic process.
The signals $d(k)$, $u(k)$, $e(k)$ and $y(k)$ denote the external input, the internal input, the error and the output, respectively.

\subsection{Structured Multiplicative Uncertainty}

Throughout this paper, given the uncertainty $\Delta$ in \eqref{eq:delta}, we make the following assumptions:

\begin{assumption}
$\left\lbrace \Delta_{i}(k) \right\rbrace$, $i=1,\ldots, m$ is a white noise with a bounded  variance $\mathbb{E}[\Delta_{i}(k)^2]\leq \sigma_i^2.$  
\label{Assp1}
\end{assumption}

\begin{assumption}
$\left\lbrace \Delta_{i}(k) \right\rbrace$, $i=1,\ldots, m$ is uncorrelated with $\left\lbrace d(k) \right\rbrace$.
\label{Assp2}
\end{assumption}

\begin{assumption}
$\left\lbrace \Delta_i(k)\right\rbrace $ and $\left\lbrace \Delta_j(k)\right\rbrace $ are correlated processes 
 $\forall i,\, j$. Denote by the vector $\eta(k)=\left[\Delta_1(k),\,\ldots,\,\Delta_m(k) \right]^{\top}$, then 
$$\mathbb{E}\left[\eta(k) \eta(k)^{\top}\right]\leq \Pi_{\Delta}.$$
\label{Assp3}
\end{assumption}
Assumptions \ref{Assp1} and \ref{Assp2} are standard in the earlier studies of classical stochastic systems and networked control systems (see, e.g., \cite{lu2002mean}).
This uncertainty may be used to model state-and input-dependent random noises in the stochastic control setting \cite{willems1976feedback,boyd1994linear}, and the communication errors and losses in networks \cite{elia2005remote,schenato2007foundations}.
The main difference from the previous studies is the existence of allowable correlations among the diagonal components in uncertainty $\Delta$, that is Assumption \ref{Assp3}. 
We aim to break a fundamental limitation with the current model and theory of networked control, that is, the communication channel noises must be independent or uncorrelated. Assumption \ref{Assp3} actually captures a wide range of channel uncertainties such as correlated noises, random delays, and transmission losses over common networks, which  could be tackled as correlated stochastic multiplicative uncertainties.

\subsection{Mean-Square Stability}
Consider the  mean-square stability problems for the interconnection in Fig.  \ref{interconnection}. 
We first provide the definition of mean-square stability from an input-output perspective as follows (see also \cite{lu2002mean,qi2017control}).

\begin{defn}
The system in Fig.\,\ref{interconnection} is said to be mean-square
stable if for any input sequence $\left\lbrace d(k)\right\rbrace $ with
bounded variance $\mathbb{E}[d(k)d(k)^{\top}] <\infty$, the variances of the
error and output sequences ${e(k)}$, ${y(k)}$ are also bounded,
i.e., $\mathbb{E}[e(k)e(k)^{\top}] <\infty$ and $\mathbb{E}[y(k)y(k)^{\top}] <\infty$.
\end{defn}
Equivalently, as to the internal stability, we mean that for any bounded initial states of the plant, the
variances of these states will converge asymptotically to the zero
matrix when $k \rightarrow \infty$. We next provide a namely generalized mean-square small-gain theorem capable of coping with correlated stochastic uncertainties,
which will play a pivotal role in our subsequent development.

\begin{theorem} 
\label{thm:diag-correlated robust}
Let $G$ be a stable LTI plant, and $\Delta$ be  a diagonally structured correlated uncertainty under \textit{Assumptions \ref{Assp1}}-\textit{\ref{Assp3}}. Then  the system in Fig. \ref{interconnection}
is mean-square stable if and only if
\begin{equation}
\rho\left(\mathrm{diag}\left(\mathrm{vec}(\Pi_{\Delta}) \right)\, \frac{1}{2\pi}\int_{-\pi}^{\pi}\hat{G}^{\ast}(e^{j\omega}) \otimes \hat{G}(e^{j\omega})\mathrm{d}\omega\right) <1.
\label{eq:small-gain-diag-correlated robust}
\end{equation}
\end{theorem}

\medskip\noindent {\em Proof.}
Sufficiency.
Define the autocorrelation matrix and the power spectral density matrix of $e(k)$ as $R_e(l)=\mathbb{E}[e(k)e(k+l)^{\top}]$
and $\Phi_e(\omega)=\sum\limits_{l=-\infty}^{\infty}R_{e}(l)e^{-j\omega l}$, then we derive that $R_e(l)=R_u(l)+R_d(l)$ and $\Phi_e(\omega)=\Phi_u(\omega)+\Phi_d(\omega)$.
In addition, $R_u(l)=\mathbb{E}[\Delta(k)R_y(l)\Delta(k+l)^{\top}]$ .

Therefore, the power spectral density matrix of the output
$y(k)$ proceeds as
\begin{equation}
\begin{aligned}
\Phi_y(\omega)&=\hat{G}(e^{j\omega})\Phi_e(\omega)\hat{G}^{\mathrm {H} }(e^{j\omega})\\
&=\hat{G}(e^{j\omega})(\Phi_u(\omega)+\Phi_d(\omega))\hat{G}^{\mathrm {H} }(e^{j\omega})\\
&=\hat{G}(e^{j\omega})R_u(0)\hat{G}^{\mathrm {H} }(e^{j\omega})+\hat{G}(e^{j\omega})R_d(0)\hat{G}^{\mathrm {H} }(e^{j\omega})\\
&=\hat{G}(e^{j\omega})\mathbb{E}[\Delta(k)R_y(0)\Delta(k)^{\top}]\hat{G}^{\mathrm {H} }(e^{j\omega})+\hat{G}(e^{j\omega})R_d(0)\hat{G}^{\mathrm {H} }(e^{j\omega}).
\nonumber
\end{aligned}
\end{equation}
Hence, the covariance matrix of the output $y(t)$ implied by $R_y(0)$ suffices that
\begin{equation}\label{eq:corance dynamics}
\begin{aligned}
&\quad\, \mathbb{E}[y(k)y(k)^{\top}]=R_y(0)=\frac{1}{2\pi}\int_{-\pi}^{\pi}\Phi_y(\omega)\mathrm{d}\omega\\
&=\frac{1}{2\pi}\int_{-\pi}^{\pi}\hat{G}(e^{j\omega})\mathbb{E}[\Delta(k)R_y(0)\Delta(k)^{\top}]\hat{G}^{\mathrm {H} }(e^{j\omega})\mathrm{d}\omega\\
&~~~~~~~~~~~+\frac{1}{2\pi}\int_{-\pi}^{\pi}\hat{G}(e^{j\omega})R_d(0)\hat{G}^{\mathrm {H} }(e^{j\omega})\mathrm{d}\omega\\
\nonumber
\end{aligned}
\end{equation}

\begin{equation}
\begin{aligned}
&~~~~~=\frac{1}{2\pi}\int_{-\pi}^{\pi}\hat{G}(e^{j\omega})\left( \mathbb{E}\left[\eta(k) \eta(k)^{\top}\right]\circ R_y(0)\right) \hat{G}^{\mathrm {H} }(e^{j\omega})\mathrm{d}\omega\\
&~~~~~~~~~~~+\frac{1}{2\pi}\int_{-\pi}^{\pi}\hat{G}(e^{j\omega})R_d(0)\hat{G}^{\mathrm {H} }(e^{j\omega})\mathrm{d}\omega.
\nonumber
\end{aligned}
\end{equation}
Denote by $\mathbb{E}\left[\eta(k) \eta(k)^{\top}\right]=\bar{\Pi}_{\Delta}$.
After the vectorization of $R_y(0)$, we rewrite that
\begin{equation}
\begin{aligned}
&\quad\, \mathrm{vec}(R_y(0))
=\frac{1}{2\pi}\int_{-\pi}^{\pi}\hat{G}^{\ast}(e^{j\omega})\otimes \hat{G}(e^{j\omega})\mathrm{d}\omega\,\mathrm{diag}\left(\mathrm{vec}(\bar{\Pi}_{\Delta}) \right)\mathrm{vec}(R_y(0))\\
&~~~~~~~~~~~~~~~~~~~~~~~~+\frac{1}{2\pi}\int_{-\pi}^{\pi}\hat{G}^{\ast}(e^{j\omega})\otimes \hat{G}(e^{j\omega})\mathrm{d}\omega\,\mathrm{vec}(R_d(0)).
\nonumber
\end{aligned}
\end{equation}
Hence, there exists a unique and finite solution for $\mathrm{vec}(R_y(0))$ if and only if $I_{n^2}-\frac{1}{2\pi}\int_{-\pi}^{\pi}\hat{G}^{\ast}(e^{j\omega})\otimes \hat{G}(e^{j\omega})\mathrm{d}\omega\,\mathrm{diag}\left(\mathrm{vec}(\bar{\Pi}_{\Delta}) \right)$ is invertible. 
A sufficient condition to guarantee this is that 
$$\rho\left(\mathrm{diag}\left(\mathrm{vec}(\bar{\Pi}_{\Delta}) \right)\frac{1}{2\pi}\int_{-\pi}^{\pi}\hat{G}^{\ast}(e^{j\omega})\otimes \hat{G}(e^{j\omega})\mathrm{d}\omega \right) <1.$$ 
Let $\mathcal{P}_n\triangleq \lbrace X\geq 0\rbrace$. Define a linear operator $\mathcal{T}_{\bar{\Pi}_{\Delta}}: \mathcal{P}_n\mapsto \mathcal{P}_n$ such that
$$\mathcal{T}_{\bar{\Pi}_{\Delta}}(X)=\frac{1}{2\pi}\int_{-\pi}^{\pi}\hat{G}(e^{j\omega})\left( \bar{\Pi}_{\Delta}\circ X\right) \hat{G}^{\mathrm {H} }(e^{j\omega})\mathrm{d}\omega.$$
Then, it follows that 
$$\rho\left(\mathrm{diag}\left(\mathrm{vec}(\bar{\Pi}_{\Delta}) \right)\frac{1}{2\pi}\int_{-\pi}^{\pi}\hat{G}^{\ast}(e^{j\omega})\otimes \hat{G}(e^{j\omega})\mathrm{d}\omega \right)=\rho(\mathcal{T}_{\bar{\Pi}_{\Delta}}).$$
Next, we need to prove $\rho(\mathcal{T}_{\bar{\Pi}_{\Delta}})\leq \rho(\mathcal{T}_{\Pi_{\Delta}})$ when $0\leq \bar{\Pi}_{\Delta}\leq \Pi_{\Delta}$. To establish this inequality, we first need three support lemmas \cite{davies2007linear,tam1989collatz}.
 
\begin{lemma}\label{lemma:schur-product}
If $A$ and $B$ are positive semi-definite (positive definite), then $A\circ B$ is positive semi-definite (positive definite). 
\end{lemma}

\begin{lemma}\label{lemma:Krein Rutman}
Given a linear operator $\mathcal{T}: \mathcal{P}_n\mapsto \mathcal{P}_n$, then $\rho(\mathcal{T})$ is an eigenvalue of
$\mathcal{T}$ together with an eigenvector $X \in \mathcal{P}_n,\,X\neq 0$, i.e. $\mathcal{T}(X)=\rho(\mathcal{T})X$.
\end{lemma}

\begin{lemma}\label{lemma:Collatz-Wielandt}
Given a linear operator $\mathcal{T}:\mathcal{P}_n\mapsto \mathcal{P}_n$, the
associated Collatz-Wielandt set is defined as $\Omega(\mathcal{T})=\left\lbrace \kappa: \exists X\in \mathcal{P}_n\setminus\lbrace 0\rbrace, \mathcal{T}(X)\geq \kappa X \right\rbrace $. Then, it holds
\begin{center}
$\sup \Omega(\mathcal{T})=\rho(\mathcal{T}).$
\end{center}
\end{lemma}
Consider
$$\begin{aligned}
&\mathcal{T}_{\Pi_{\Delta}}(X)-\mathcal{T}_{\bar{\Pi}_{\Delta}}(X)=\frac{1}{2\pi}\int_{-\pi}^{\pi}\hat{G}(e^{j\omega})\left( (\Pi_{\Delta}-\bar{\Pi}_{\Delta})\circ X\right) \hat{G}^{\mathrm {H} }(e^{j\omega})\mathrm{d}\omega .
\end{aligned}$$
Lemma \ref{lemma:schur-product} and $\Pi_{\Delta}-\bar{\Pi}_{\Delta}\leq 0$ yield that $ (\Pi_{\Delta}-\bar{\Pi}_{\Delta})\circ X\geq 0, \forall X\geq 0$, then 
$\mathcal{T}_{\Pi_{\Delta}}(X)-\mathcal{T}_{\bar{\Pi}_{\Delta}}(X)\geq 0.$ Inspired by invoking Lemma \ref{lemma:Krein Rutman}, there must exist $X_{\rho}\in \mathcal{P}_n$
as the eigenvector  such that $\mathcal{T}_{\bar{\Pi}_{\Delta}}(X_\rho)=\rho(\mathcal{T}_{\bar{\Pi}_{\Delta}})X_\rho.$ Therefore, it follows that 
$\rho(\mathcal{T}_{\bar{\Pi}_{\Delta}})X_\rho\leq \mathcal{T}_{\Pi_{\Delta}}(X_\rho)$ and, from Lemma \ref{lemma:Collatz-Wielandt}, $\rho(\mathcal{T}_{\bar{\Pi}_{\Delta}})\in \Omega(\mathcal{T}_{\Pi_{\Delta}})$. Hence, $\rho(\mathcal{T}_{\bar{\Pi}_{\Delta}})\leq \sup \Omega(\mathcal{T}_{\Pi_{\Delta}})=\rho(\mathcal{T}_{\Pi_{\Delta}})$. Finally, a sufficient condition to guarantee the mean-square stability is $\rho(\mathcal{T}_{\Pi_{\Delta}})<1$, so is to \eqref{eq:small-gain-diag-correlated robust}.

Necessity. 
Assume that there exists the uncertainty $\hat{\Delta}$ with $\bar{\Pi}_{\hat{\Delta}}\leq \Pi_{\hat{\Delta}}$
such that $\rho\left( \mathcal{T}_{\Pi_{\hat{\Delta}}})\right) \geq 1$.  Then by the continuity and the monotonicity proved above,  we can always find a uncertainty $\hat{\tilde{\Delta}}$ with $\bar{\Pi}_{\hat{\tilde{\Delta}}}\leq \Pi_{\hat{\Delta}}$ satisfying $\rho(\mathcal{T}_{\bar{\Pi}_{\hat{\tilde{\Delta}}}})=1$,
which implies the non-uniqueness of $R_y(0)$, and
then  leads to a contradiction. 
Thus, $\rho(\mathcal{T}_{\Pi_{\Delta}})<1$ is also necessary, so is to \eqref{eq:small-gain-diag-correlated robust}.
\hfill $\blacksquare$

\noindent
Consider also the uncorrelated uncertainty as a special case.
\begin{assumption}
$\left\lbrace \Delta_i(k)\right\rbrace $ and $\left\lbrace \Delta_j(k)\right\rbrace $ are uncorrelated processes 
 $\forall i,\, j$, i.e.,
\begin{center}
$\mathbb{E}[\Delta_i(k)\Delta_j(k)]=0,\; \forall\,i\neq j.$
\end{center}
\label{Assp4}
\end{assumption}
The following result, herein referred to as the mean-square
small-gain theorem, is adapted from \cite{lu2002mean} (see also \cite{qi2017control} and \cite{elia2005remote}
), which provides a necessary and sufficient condition
for mean-square  stability subject to uncorrelated uncertainties.

\begin{coro} 
\label{lm:MSS1}
Let $G$ be a stable LTI plant, and $\Delta$ be a diagonally structured uncorrelated uncertainty under Assumptions \ref{Assp1}-\ref{Assp2} and \ref{Assp4}. Then the system in Fig. \ref{interconnection}  
is mean-square stable if and only if
\begin{equation}
\rho\left(\mathrm{diag}\left( \sigma_1^2,\cdots, \sigma_m^2\right) \, H\right) <1,
\label{eq:MSS1}
\end{equation}
where 
\begin{equation}
H=\begin{bmatrix}
\norm{G_{11}}_2^2&\cdots&\norm{G_{1m}}_2^2\\
\vdots&\ddots&\vdots\\
\norm{G_{m1}}_2^2&\cdots&\norm{G_{mm}}_2^2
\end{bmatrix}.
\label{eq:G}
\end{equation}
\end{coro}
It is worth noting that Theorem \ref{thm:diag-correlated robust} can reduce to the result above if the uncertainties change to uncorrelated. We omit the details herein.

\section{Main Results} 
Our ultimate goal is developing mean-square stabilization conditions for general LTI plants using the obtained generalized mean-square small-gain theorem as a coherent technical approach.  Nevertheless, compared with the existing results concentrated on the uncorrelated uncertainties (see for example \cite{chen2012lqg} and \cite{qi2017control}), the mean-square stabilizability via output feedback under  correlated uncertainties proves fundamentally more difficult, which appears to be a nontrivial task and is currently unavailable.
In this paper, two low-order systems are chosen serving as a case study. Firstly, a  SISO first-order unstable plant has been put into consideration and the communication channels between the plant and the controller are perturbed by multiplicative uncertainties. Next, we turn to resolve the mean-square consensusability problem of the two-agent system. Each agent admits a first-order unstable dynamics. We can also observe that one agent communicates with another through a communication channel perturbed by a multiplicative uncertainty.

\subsection{Multiplicative $\&$ Divisive Uncertainties}

\begin{figure}
\begin{center}
\includegraphics[width=9.8cm, height=4.3cm]{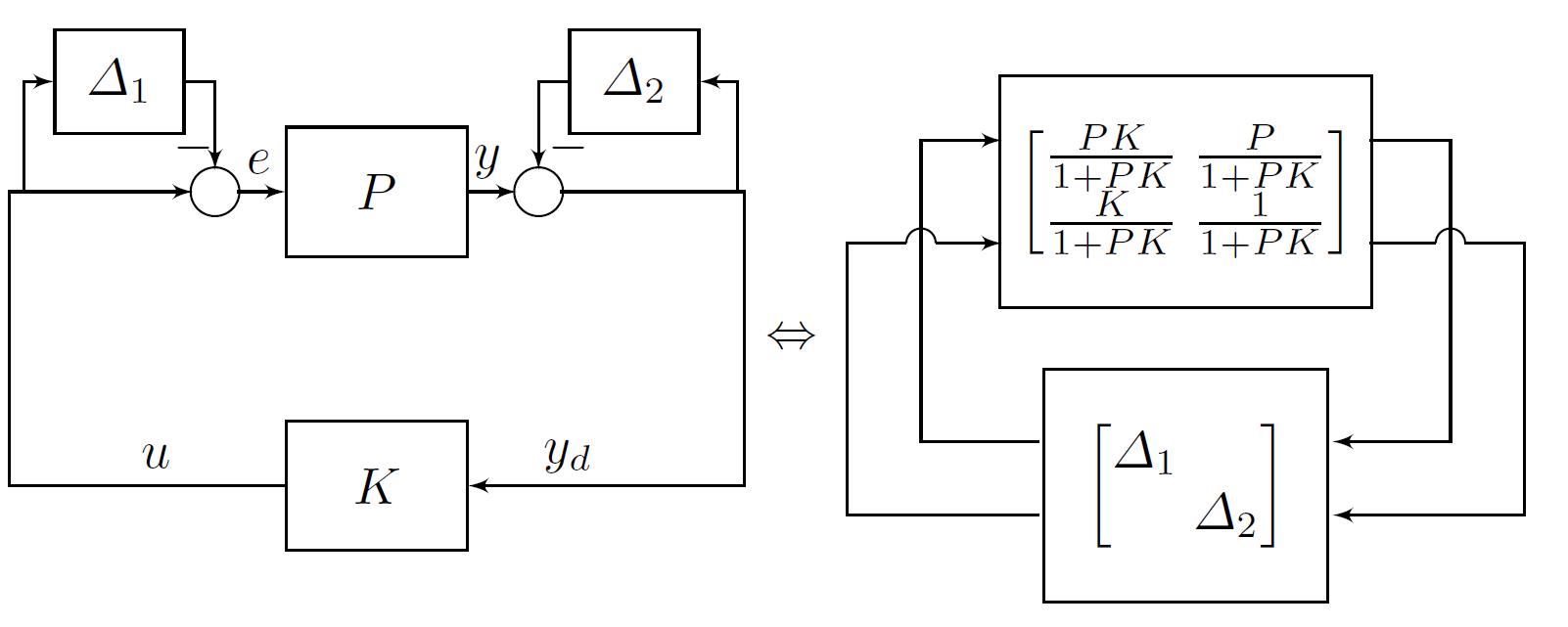}
\end{center}

\caption{Reshaped loop of system with the correlated uncertainties at plant input and output.}
\label{reshaped-siso}
\end{figure}

We focus on the uncertain system depicted in Fig. \ref{reshaped-siso}. The nominal plant $P$ represents a SISO first-order system  with relative degree $\tau=1$ such that
\begin{equation}
P(z)=\dfrac{1}{z-p_1},~~~~~|p_1|>1,
\label{eq:first-order unstable case 1}
\end{equation}
and  $\tau=0$ such that
\begin{equation}
~~~~~P(z)=\dfrac{z-s_1}{z-p_1},~~~~~|s_1|>1,~|p_1|>1.
\label{eq:first-order unstable case 2}
\end{equation}
The plants $P(z)$ in \eqref{eq:first-order unstable case 1} and \eqref{eq:first-order unstable case 2} both admit an unstable pole $p_1$, whereas the latter contains one nonminimum phase zero $s_1$.  The uncertainties $\Delta_1$ and $\Delta_2$ are usually named as the multiplicative and divisive uncertainties,  both of which are assumed to be zero-mean stochastic processes. 

\noindent
After a linear fractional transformation, we obtain a $G-\Delta$ loop at the right hand side in Fig. \ref{reshaped-siso}, where 
\begin{equation}
G=
\begin{bmatrix}
\dfrac{PK}{1+PK}&\dfrac{P}{1+PK}\\
\dfrac{K}{1+PK}&\dfrac{1}{1+PK}
\end{bmatrix}
\end{equation}
and the uncertainty
$\Delta=\mathrm{diag}(\Delta_1,\Delta_2)$ satisfies Assumption \ref{Assp1}-\ref{Assp3} with 
\begin{equation}\label{eq:Pi}
\Pi_{\Delta}=\begin{bmatrix}
\sigma_1^2&\sigma_{12}\\
\sigma_{12}&\sigma_2^2
\end{bmatrix}.
\end{equation}
The following results show that the mean-square stabilizability conditions of the first-order plant under consideration.

\begin{theorem}\label{thm:siso two block}
(i) For $P(z)$  in \eqref{eq:first-order unstable case 1},  the system in  Fig. \ref{reshaped-siso} is mean-square stabilizable if and only if 
\begin{equation}\label{eq:first-order iff condition}
(\sigma_1^2-2\sigma_{12})(p_1^2-1)+\sigma_2^2p_1^2<1.
\end{equation}

(ii) For $P(z)$ in \eqref{eq:first-order unstable case 2},  the system in  Fig. \ref{reshaped-siso} is mean-square stabilizable if
 \begin{equation}\label{eq:first-order iff condition2}
(\sigma_1^2-2\sigma_{12}+\sigma_2^2)\frac{p_1^2s_1^2}{(p_1-s_1)^2}+\frac{\sigma_1^2p_1^2-2\sigma_{12}s_1p_1+\sigma_2^2s_1^2}{(p_1-s_1)^2}<1.
\end{equation}
\end{theorem}

\medskip\noindent {\em Proof.}
In light of Theorem \ref{thm:diag-correlated robust}, the system in  Fig. \ref{reshaped-siso} is mean-square stabilizable if and only if 
$\inf_{K:~G~\mathrm{is~stable}}
\rho\left(\mathrm{diag}\left(\mathrm{vec}(\Pi_{\Delta}) \right)\, \frac{1}{2\pi}\int_{-\pi}^{\pi}\hat{G}^{\ast}(e^{j\omega}) \otimes \hat{G}(e^{j\omega})\mathrm{d}\omega\right) \! \!< \!1.$
Specifically,  it follows that
\begin{small}
\begin{equation}
\begin{aligned}
&\!\! \inf_{K:~G~\mathrm{is~stable}}
\!\! \!\! \!\! \rho\left(\mathrm{diag}\left(\mathrm{vec}(\Pi_{\Delta}) \right)\ \frac{1}{2\pi}\int_{-\pi}^{\pi} \left| \frac{1}{1+P(e^{j\omega})K(e^{j\omega})} \right|^{2}\right. 
\\
&~~~~~~~~~~~~~~~~~~~~~~\left. \left( 
\begin{bmatrix}\left| P(e^{j\omega})\right|^{2}\\P^{\ast}(e^{j\omega})\\P(e^{j\omega})\\1
\end{bmatrix}\begin{bmatrix}
\left| K(e^{j\omega})\right|^{2}&K^{\ast}(e^{j\omega})&K(e^{j\omega})&1
\end{bmatrix} 
\right) \mathrm{d}\omega \right)  \! \!< \!1.
\end{aligned}
\label{eq:SISO correlated}
\nonumber
\end{equation}
\end{small}
Toward the first-order system, it is rather enough to consider only the constant output feedback, i.e.,
 $K=K(e^{j\omega})$.   We then have
 \begin{small} 
\begin{equation}
\begin{aligned}
\!\! \inf_{K:~G~\mathrm{is~stable}}
\!\! \!\!  &\rho\left(\mathrm{diag}\left(\mathrm{vec}(\Pi_{\Delta}) \right)\
 \frac{1}{2\pi}\int_{-\pi}^{\pi} \left| \frac{1}{1+P(e^{j\omega})K} \right|^{2}
 \begin{bmatrix}\vert P(e^{j\omega})\vert^{2}\\P^{\ast}(e^{j\omega})\\P(e^{j\omega})\\1
\end{bmatrix}\mathrm{d}\omega
\begin{bmatrix}
K^{2}&K&K&1
\end{bmatrix} 
 \right)   \! \!< \!1.
\end{aligned}
\nonumber
\label{eq:SISO correlated2}
\end{equation}
\end{small}
Apparently,  the matrix inside the spectral radius operator is rank-one,  which,  along with Lemma \ref{lemma:Krein Rutman}, indicates the equality between the spectral radius  and the trace of the given matrix. The mean-square stabilizable condition reduces to
\begin{small}
\begin{equation}
\begin{aligned}
&\!\! \inf_{K:~G~\mathrm{is~stable}}
\frac{1}{2\pi}\int_{-\pi}^{\pi}\left| \frac{1}{1+P(e^{j\omega})K} \right|^{2}\times
\\
&\left(\sigma_1^2\vert P(e^{j\omega})\vert^{2}K^2+\sigma_{12}(P^{\ast}(e^{j\omega})+P(e^{j\omega}))K+\sigma_2^2 \right)\mathrm{d}\omega <1.
\end{aligned}
\nonumber
\end{equation}
\end{small}

\noindent \textbf{Case I: }$P(e^{j\omega})=1/(e^{j\omega}-p_1)$.  It is easy to verify  that $-1+p_1<K<1+p_1$.
Denote by 
\begin{small}
$$
\begin{aligned}
f_{\mathrm{I}}(K)&=\frac{1}{2\pi}\int_{-\pi}^{\pi}\left| \frac{1}{1+\frac{1}{e^{j\omega}-p_1}K} \right|^{2}\times
\\
&~~~~~~~~~~~\left(\sigma_1^2\vert \frac{1}{e^{j\omega}-p_1}\vert^{2}K^2+\sigma_{12}(\frac{1}{e^{-j\omega}-p_1}+\frac{1}{e^{j\omega}-p_1})K+\sigma_2^2 \right)\mathrm{d}\omega
\\
&=\frac{1}{1-(p_1-K)^2}\left( (\sigma_1^2-2\sigma_{12})K^2+2\sigma_2^2p_1K+\sigma_2^2(1-p_1^2)\right).
\end{aligned}
$$ 
\end{small}
Taking the derivative of  $f_{\mathrm{I}}(K)$ with $K$, 
we have
$$
 f_{\mathrm{I}}'(K)=\frac{2(\sigma_1^2-2\sigma_{12}+\sigma_2^2)}{(1-(p_1-K)^2)^2}\left( p_1k^2+(1-p_1^2)k\right) .$$
Then we obtain  $K^\ast=p_1-1/p_1$ such that  $ f_{\mathrm{I}}'(K^\ast)=0$.
Note that $\sigma_1^2-2\sigma_{12}+\sigma_2^2>0$.  
It  is evident that $f_{\mathrm{I}}'(K)<0$  when  $-1+p_1<K<K^\ast$ and 
$f_{\mathrm{I}}'(K)>0$ when $K^\ast<K<1+p_1$.
Consequently,  it can be established that
$$\inf_{K}  f_{\mathrm{I}}(K)=f_{\mathrm{I}}(K^\ast)=(\sigma_1^2-2\sigma_{12})(p_1^2-1)+\sigma_2^2p_1^2,$$
which in turn implies the mean square stabilizability,  provided
that the condition  \eqref{eq:first-order iff condition} holds.

\noindent \textbf{Case II: }$P(e^{j\omega})=(e^{j\omega}-s_1)/(e^{j\omega}-p_1)$.   Using Jury stability criterion \cite{ogata1994discrete},   we first confirm the range of $K$ with
\begin{small}
$$\min\lbrace -\frac{1-p_1}{1-s_1},~-\frac{1+p_1}{1+s_1}\rbrace<K<\max\lbrace -\frac{1-p_1}{1-s_1},~-\frac{1+p_1}{1+s_1}\rbrace$$
\end{small}
 such that the close-loop system $G$ is stable.
Next,  we consider
\begin{small}
$$
\begin{aligned}
&f_{\mathrm{II}}(K)=\frac{1}{2\pi}\int_{-\pi}^{\pi}\left| \frac{1}{1+\frac{e^{j\omega}-s_1}{e^{j\omega}-p_1}K} \right|^{2}\times
\\
&~~~~~~~~~\left(\sigma_1^2\vert \frac{e^{j\omega}-s_1}{e^{j\omega}-p_1}\vert^{2}K^2+\sigma_{12}(\frac{e^{-j\omega}-s_1}{e^{-j\omega}-p_1}+\frac{e^{j\omega}-s_1}{e^{j\omega}-p_1})K+\sigma_2^2 \right)\mathrm{d}\omega.
\end{aligned}
$$ 
\end{small}
Let $x=(p_1+s_1K)/(1+K)\in (-1,1)$ be the stable pole of the close-loop system.
We obtain $f_{\mathrm{II}}(K)=g(x),$ where
\begin{small}
\begin{equation}\label{eq:g(x)}
\begin{aligned}
g(x)&=\sigma_1^2\frac{(x-p_1)^2}{(p_1-s_1)^2}\frac{1+s_1^2-2s_1x}{1-x^2}-2\sigma_{12}\frac{(x-p_1)(x-s_1)}{(p_1-s_1)^2}\frac{1+s_1p_1-(s_1+p_1)x}{1-x^2}\\
&~~~~~~~~~~+\sigma_2^2\frac{(x-s_1)^2}{(p_1-s_1)^2}\frac{1+p_1^2-2p_1x}{1-x^2}.
\end{aligned}
\nonumber
\end{equation}
\end{small}
The condition in \eqref{eq:first-order iff condition2} hence can be established through letting $g(0)<1$.
\hfill $\blacksquare$

\noindent
Theorem \ref{thm:siso two block} provides a complete solution to the mean-square
stabilizability problem in the first-order case against the multiplicative and divisive stochastic uncertainties simultaneously.
It is important to note that for the minimum phase plant \eqref{eq:first-order unstable case 1}, 
Theorem \ref{thm:siso two block},  \eqref{eq:first-order iff condition}  shows that
the mean-square stabilizability condition becomes proportionally
more demanding as the variances $\sigma_1^2,~\sigma_2^2$ increase. Also, the distance between the unstable pole $p_1$ and the unit circle, as a measure
of the system's instability, plays a central role
in the mean-square stabilization. As to the nonminimum phase plant  \eqref{eq:first-order unstable case 2}, it is clear from \eqref{eq:first-order iff condition2}
that the unstable pole $p_1$ and nonminimum phase zero $s_1$ coupled together codetermine the mean-square
stabilizability condition. Especially when the pole and zero are getting close to each other, it is rather difficult to satisfy the  stabilization condition.
Note that the influence of the correlation pattern $\sigma_{12}$ in stabilization has been demonstrated clearly in \eqref{eq:first-order iff condition} and 
\eqref{eq:first-order iff condition2}. In addition, for the minimum phase plant, if $\sigma_{12}=\sigma_2^2=0$, the condition \eqref{eq:first-order iff condition} reduces
to $\sigma_1^2(p_1^2-1)<1.$ If $\sigma_1^2=\sigma_{12}=0$, we obtain the reduced condition $\sigma_2^2p_1^2<1$.  Those two reduced conditions coincide with the existing results in \cite{chen2012lqg,braslavsky2007feedback}.

\subsection{Two-Agent System}
\begin{figure}
\begin{center}
\begin{tikzpicture}[auto, node distance=3cm,>=latex']
\tikzstyle{sum} = [draw, circle, node distance=3cm]
  \node [sum] (sum1) at (1,0) {$1$};
  \node [sum] (sum2) at (4,0) {$2$};
           
  \draw [->] (1.4,0.2) -- (3.6,0.2);
  \draw [<-] (1.4,-0.2) -- (3.6,-0.2);
\end{tikzpicture}
\end{center}
\caption{The two-agent system.}
\label{twoagent}
\end{figure}
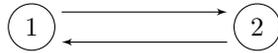

In this subsection, we examine the simplest multi-agent system as a case study, that is a two-agent system (Fig. \ref{twoagent}).

\begin{figure}
\begin{center}
\includegraphics[width=9.8cm, height=4.3cm]{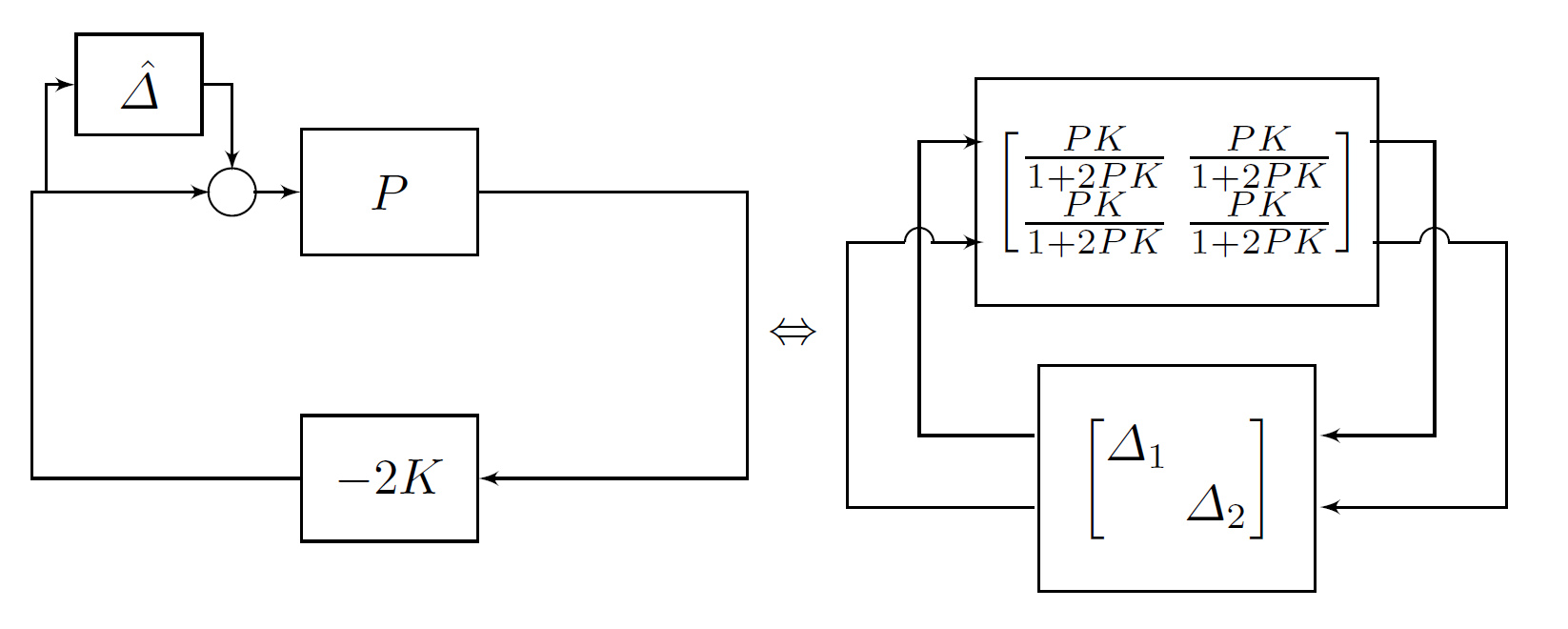}
\end{center}

\caption{Reshaped loop of two-agent system with correlated uncertainties.}
\label{reshaped-two-agent}
\end{figure}

\noindent
The dynamics are described by
\begin{equation}\label{eq:two agent system}
\begin{aligned}
x_1(k+1)&=p_1x_1(k)+u_1(k)
\\
x_2(k+1)&=p_1x_2(k)+u_2(k)
\end{aligned}
\end{equation}
with $|p_1|\geq 1$.
The consensus protocol is given as follows:
\begin{equation}\label{eq:consensus}
\begin{aligned}
u_1(k)&=(1+\Delta_1(k))K(x_2(k)-x_1(k))
\\
u_2(k)&=(1+\Delta_2(k))K(x_1(k)-x_2(k)),
\end{aligned}
\end{equation}
where $K$ is the feedback gain.
Clearly, the agents are perturbed by stochastic uncertainties in their control input channels.
We then focus on the robust consensus problem
in the mean square sense. Here we say that a group of
agents achieve\textit{ mean square consensus} if the states of the
agents converge asymptotically to a common state under
the mean square criterion, i.e., $\lim_{k\rightarrow \infty} \mathbb{E}[x_1(k)-x_2(k)]=0.$
Consider the error dynamics
\begin{equation}
\label{eq:error}
\begin{aligned}
e(k+1)&=p_1e(k)-(2+\Delta_1(k)+\Delta_2(k))Ke(k)
\\
&=p_1e(k)-2(1+\hat{\Delta}(k))Ke(k),
\end{aligned}
\end{equation}
where $e(k)=x_1(k)-x_2(k)$ and 
$$\hat{\Delta}(k)=\frac{1}{2}\textbf{1}_2^{\top}\begin{bmatrix}
\Delta_1(k)&\\&\Delta_2(k)
\end{bmatrix}\textbf{1}_2.
$$
Denote by $P(z)=\frac{1}{z-p_1}.$  The error dynamics \eqref{eq:error} can be transferred into the framework of $G-\Delta$ loop equivalently, as depicted in 
Fig. \ref{reshaped-two-agent}.  The uncertainties $\Delta_1$ and $\Delta_2$ are assumed to satisfy Assumptions \ref{Assp1}-\ref{Assp3} with $\Pi_{\Delta}$ in \eqref{eq:Pi}.

\begin{theorem}\label{thm:siso two agent}
There exists a consensus protocol \eqref{eq:consensus} achieving
mean square consensus for the two-agent system in \eqref{eq:two agent system} if and only if 
\begin{equation}\label{eq:two-agent iff condition}
\frac{1}{4}(\sigma_1^2+2\sigma_{12}+\sigma_2^2)(p_1^2-1)<1.
\end{equation}

\end{theorem}

\noindent
Following similar steps in the proof of Theorem \ref{thm:siso two block}, it is
not difficult to verify the condition \eqref{eq:two-agent iff condition}. The details are omitted here for conciseness.
As a remark, we point out that 
when the agents contain no eigenvalues outside the unit circle, i.e., $p_1 = \pm 1$, one can show that
the condition \eqref{eq:two-agent iff condition} always holds. 
The implication then is that  no matter how large the variance  of the uncertainty is, we can always find a
feedback $K$ to guarantee the robust mean square consensus.

\section{Illustrative Example}
We use two examples to examine the preceding results in Theorem \ref{thm:siso two block} and \ref{thm:siso two agent}.

\begin{figure}[h]
\centering
\begin{minipage}{.48\textwidth}
  \centering
  \includegraphics[width=\textwidth]{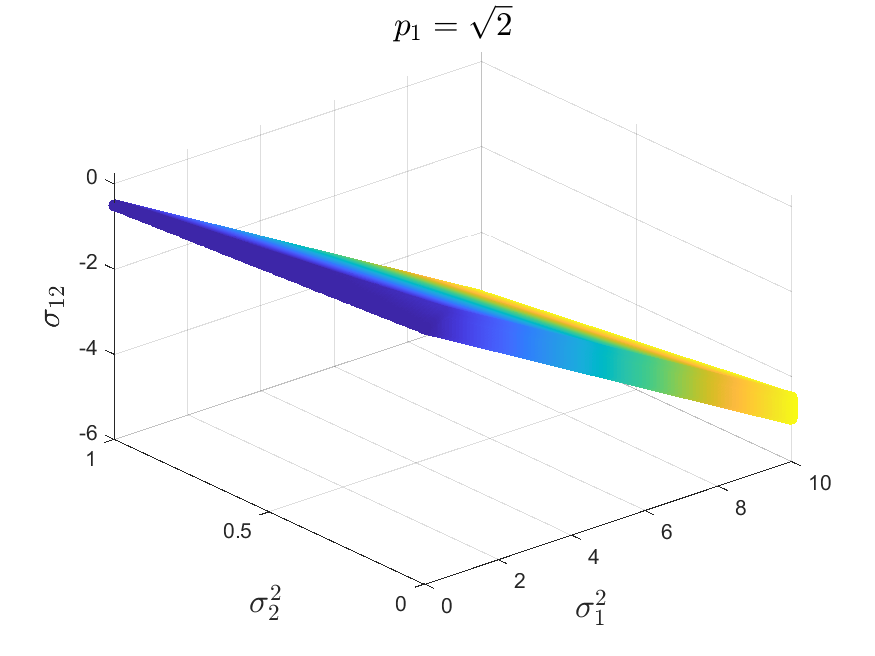}
$(a)$
\end{minipage}
\hfill
\begin{minipage}{.48\textwidth}
  \centering
  \includegraphics[width=\textwidth]{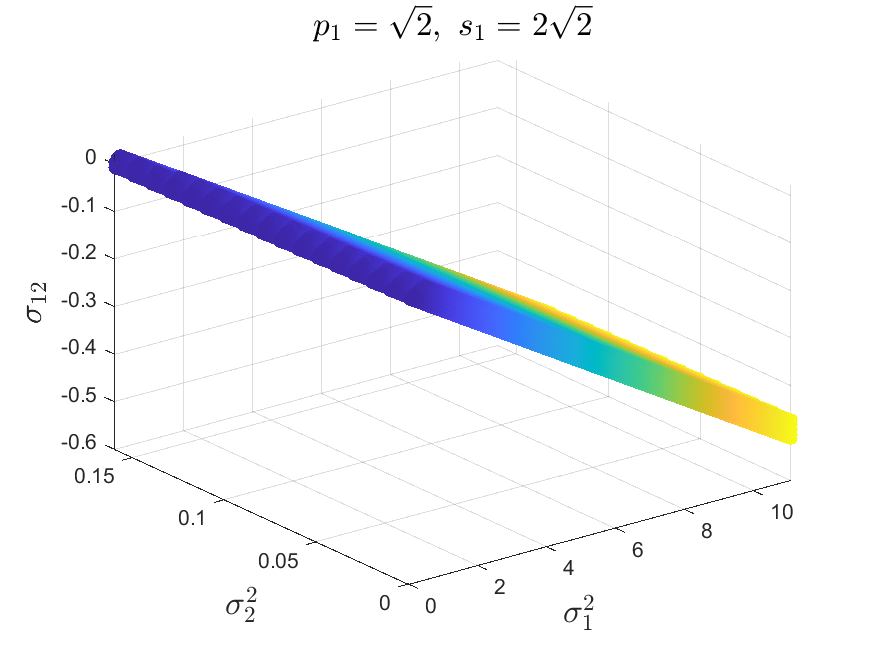}
  $(b)$
\end{minipage}
\caption{Mean-square stabilizable region for $(\sigma_1^2,\sigma_2^2,\sigma_{12})$.}
\label{fig:mss-region-uncertainties}
\end{figure}

\textbf{Example 1}
Consider first the first-order system \eqref{eq:first-order unstable case 1} with an unstable pole $p_1=\sqrt{2}$. 
It is  instructive to examine the three-dimensional (3-D) manifold of $(\sigma_1^2,\sigma_2^2,\sigma_{12})$, to see the 
corresponding mean-square stabilizable region (also mean-square unstabilizable region) with respect to the stochastic uncertainties $\Delta_1$ and $\Delta_2$.
Fig. \ref{fig:mss-region-uncertainties} $(a)$ gives approximately the mean-square stabilizable region in terms of $(\sigma_1^2,\sigma_2^2,\sigma_{12})$, which constitutes an unbounded
convex hull. We next examine the the first-order nonminimum phase system \eqref{eq:first-order unstable case 2} with a nonminimum phase zero $s_1=2\sqrt{2}$. Fig. \ref{fig:mss-region-uncertainties} $(b)$ characterizes  the
 mean-square stabilizable region accordingly. Apparently, the presence of nonminimum phase zero drastically reduces the feasible region for the mean-square stabilization, following 
one's long-held intuition.

\begin{figure}[h!]
\centering
\begin{minipage}{.48\textwidth}
  \centering
  \includegraphics[width=\textwidth]{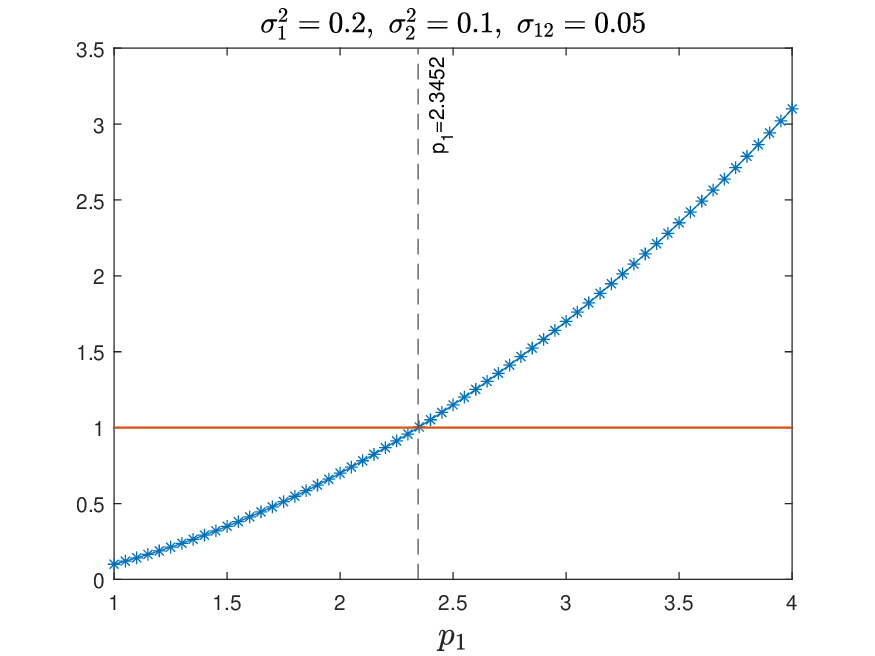}
$(a)$
\end{minipage}
\hfill
\begin{minipage}{.48\textwidth}
  \centering
  \includegraphics[width=\textwidth]{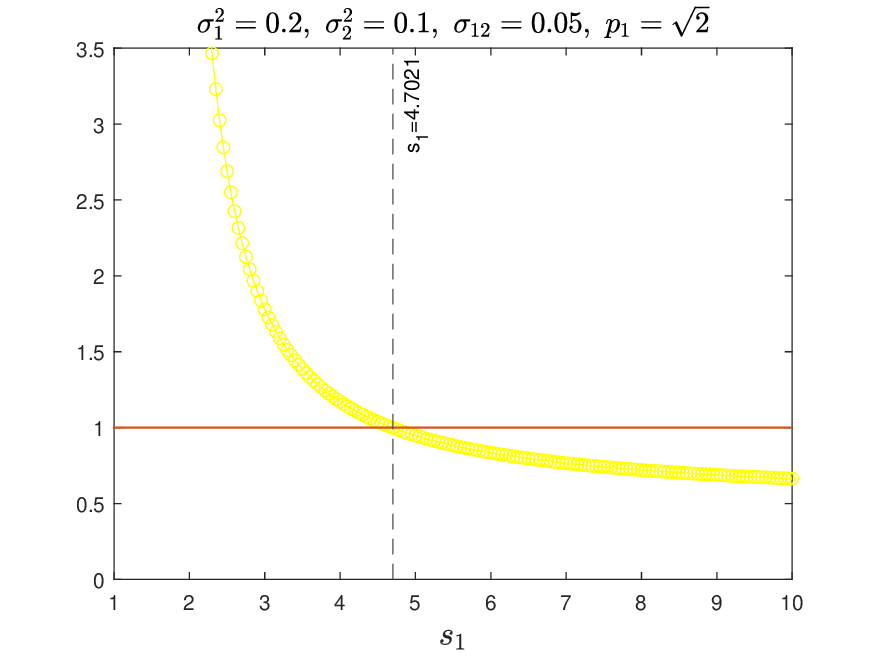}
$(b)$
\end{minipage}
\caption{Mean-square stabilizable region for $p_1$ and $s_1$.}
\label{fig:mss-region-p1}
\end{figure}

We may also fix the triple $(\sigma_1^2,\sigma_2^2,\sigma_{12})=(0.2,0.1,0.05)$ and verify the mean-square  stabilizable region for all $p_1\geq 1$. Fig. \ref{fig:mss-region-p1} $(a)$
shows that when $p_1\in [1,~2.3453)$, the system \eqref{eq:first-order unstable case 1} is mean-square stabilizable. Next, with fixing $p_1=\sqrt{2}$, we can conclude from 
Fig. \ref{fig:mss-region-p1} $(b)$ that the nonminimum phase   system \eqref{eq:first-order unstable case 2} is mean-square stabilizable if $s_1\geq 4.7021$. This observation
is consistent with a  well-accepted agreement, that is a zero-pole closeness is detrimental to the robust stabilization.

\begin{figure}[h!]
\begin{center}
\includegraphics[width=9.8cm, height=6.5cm]{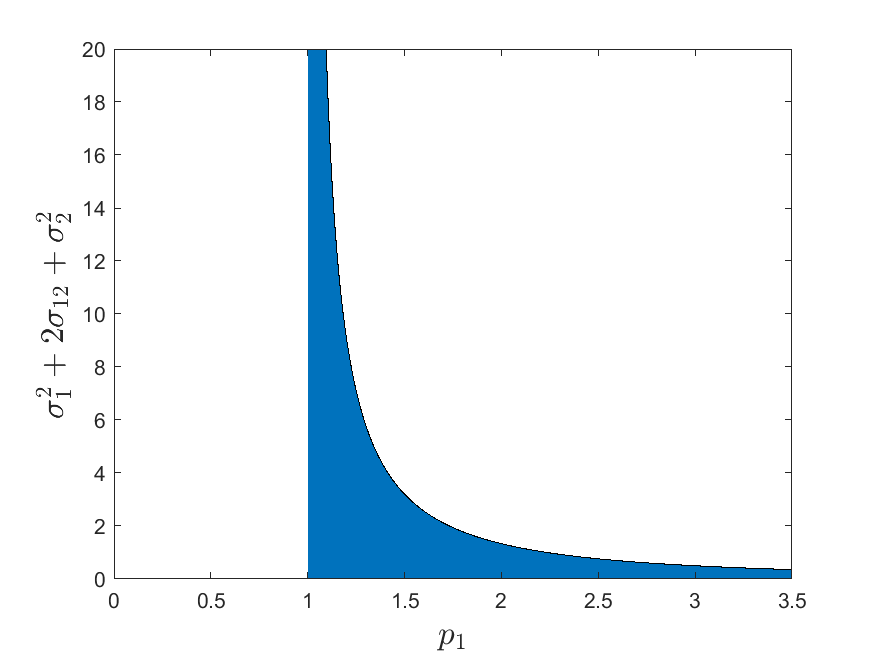}
\end{center}
\caption{The region of mean-square consensusability for $p_1$ and  $\sigma_1^2+2\sigma_{12}+\sigma_2^2$}
\label{fig:two-agent-region}
\end{figure}

\textbf{Example 2}
Following from \eqref{eq:two-agent iff condition}, the mean-square consensusability of the two-agent system \eqref{eq:two agent system} is codetermined by the unstable pole $p_1$ and the variance $\sigma_1^2+2\sigma_{12}+\sigma_2^2$ together and the feasible region has been filled in Fig. \ref{fig:two-agent-region}. The inversely proportional relationship between  $p_1$ and $\sigma_1^2+2\sigma_{12}+\sigma_2^2$ hence is clearly.

\section{Conclusion}
In this paper we have studied mean-square stabilizability and consensusability problems for networked control systems over uncertain communication channels. 
We modelled each communication channel as an ideal transmission system subject to a multiplicative stochastic perturbation. All perturbations from different channels are allowed to
be correlated in spatial.
We first presented fundamental conditions to ensure the mean-square stability of the open-loop stable system under such uncertainties.
Next, given two kind of unstable low-order systems, we provided the stabilizability or  consensusability  conditions linking together such system dynamics and   uncertainty variances, under which the robust stable or consensus can be ensured. 
Notably, all obtained main conditions are necessary and sufficient in the paper.

\bibliographystyle{IEEEtran}

\bibliography{mycite}

\end{document}